\pdfoutput=1
\documentclass{article}
\usepackage{graphicx, indentfirst, hyperref}
\usepackage[a4paper, margin = 3cm]{geometry}
\usepackage[numbers]{natbib}
\makeatletter
\let\mailmark\@fnsymbol
\makeatother

\newcommand*{\cf}{\emph{cf.}}
\newcommand*{\eg}{\emph{eg.}}
\newcommand*{\ie}{\emph{i.e.}}
\newcommand*{\vs}{\emph{vs.}}

\newcommand*{\lb}{\linebreak[1]}
\newcommand*{\prog}[1]{\emph{#1}}
\newcommand*{\figref}[1]{Figure \ref{fig:#1}}
\newcommand*{\secref}[1]{Section \ref{sec:#1}}
\newcommand*{\ssref}[1]{Subsection \ref{ssec:#1}}
\let\thxmark\textsuperscript
\begin{document}
\title{%
	Software refactoring and rewriting:\\
	from the perspective of code transformations%
}
\author{Yu Liu\thxmark{1,\mailmark{1}}}
\date{}
\maketitle
\begingroup
\renewcommand{\thefootnote}{\fnsymbol{footnote}}
\footnotetext[1]{\ Correspondence e-mail: \texttt{liuyu91@ihep.ac.cn}.}
\endgroup
\footnotetext[1]{\ %
	Institute of High Energy Physics, Chinese Academy of Sciences,
	Beijing 100049, People's Republic of China.%
}

\section*{Abstract}

Keywords: software refactoring, code transformation,
code equivalence, semiformal method.

To refactor already working code while keeping reliability, compatibility
and perhaps security, we can borrow ideas from micropass/nanopass compilers.
By treating the procedure of software refactoring as composing code
transformations, and compressing repetitive transformations with automation
tools, we can often obtain representations of refactoring processes short
enough that their correctness can be analysed manually.  Unlike in
compilers, in refactoring we usually only need to consider the codebase
in question, so regular text processing can be extensively used, fully
exploiting patterns only present in the codebase.  Aside from the direct
application of code transformations from compilers, many other kinds of
equivalence properties may also be exploited.  In this paper, two refactoring
projects are given as the main examples, where 10--100 times simplification has
been achieved with the application of a few kinds of useful transformations.

\section{Introduction}

Refactoring is hard.  Refactoring already working code is harder, even if the
product can be obviously much simpler, because of the need to avoid hurting
reliability, compatibility or even security (if the proposed change is made by
a newcomer).  To deal with this problem, we can learn from micropass/nanopass
compilers \cite{keep2012}, which focus on representing the process as small
\emph{compilation passes} that can be analysed individually.  By treating
the procedure of refactoring as composing \emph{code transformations},
and compressing repetitive transformations with automation tools
(\cf\ \cite{bentley1986} for a classic example), we can often obtain
representations (\emph{code transformers}) of non-trivial code
transformations short enough that they can be analysed manually.

Unlike compilers which are normally expected to produce correct outputs for all
valid inputs, refactoring is usually done on a one-of-a-kind basis, with the
input predetermined (or nearly so, if the codebase also evolves upstream during
the procedure; we will also mention scenarios where similar transformations
need to be applied to multiple similar codebases in \ssref{discuss}).
Therefore in contrast to compilers which usually need structural
intermediate representations (IRs) like S-expressions for comfortable
encoding of transformations, in refactoring we can fully exploit patterns
only present in the codebase in question, thus avoiding the big overhead
in implementing conversions from/to these IRs (\cf\ \ssref{xsp3-methods};
nevertheless in \ssref{int-repr}, we will also show a way IRs, whether
structural or not, can help to ensure the correctness of refactoring
transformations).  In refactoring, we often do transformations selectively
and sometimes do inverse transformations: the former can be seen throughout
this paper, and an example for the latter can be seen in \ssref{discuss}.

In real applications, it is usually impractical to formally prove the
correctness of a codebase; instead, the programmer's experience is relied
upon, often in combination with some kind of semiformal reasoning; the
same problem exists with refactoring, and the same solution can be applied.
In the case of refactoring, apart from the experience and reasoning commonly
used in regular programming, it can also be instructional to borrow ideas
from code transformations in compilers.  Aside from direct application
of transformations like the dead-code elimination (\cf\ Subsection
\ref{ssec:syno-cjdns} and \ref{ssec:syno-xsp3}), the implicit notion of
\emph{equivalent code}, stemming from the idea that many transformations
convert their input languages into equivalent but more primitive
output languages, is followed throughout this paper.

Our transformation-based approach has been successfully applied
in more than 10 internal or open-source refactoring projects, quite a few
of them with sufficiently big changes that they may be properly called
rewriting projects.  In the following sections, we will demonstrate the
transformation-based approach to refactoring/rewriting mainly with two
examples where literally 10--100 times simplification has been achieved,
and the resulting products have been tested in real-world use.  Considering
the diversity in possible refactoring/rewriting requirements, we do not
attempt to establish a very systematic treatment of the topic, but
we wish this paper could become a motivation for such treatments.

\section{Refactoring the build system of \prog{cjdns}}\label{sec:cjdns}
\subsection{Background}\label{ssec:bg-cjdns}

\prog{cjdns} (\url{https://github.com/cjdelisle/cjdns}) is a mesh-networking
scheme with builtin end-to-end encryption and mandatory association between
network addresses and public keys used in said encryption.  We used it to
facilitate the access to a group of about 13 computers, with differing
hardware/software configurations for various purposes, separated in a few
locations with regular consumer-grade network access.  Before this deployment
actually took place, the build system of \prog{cjdns} raised our attention, and
its refactoring is the subject of this section; the deployment was done after
our refactoring, and although we had issues with \prog{cjdns} (which lead to our
migration to other technologies), during our 8-month use of \prog{cjdns} none
of the issues were attributed to the refactoring.  Our refactoring can be
accessed at \url{https://github.com/CasperVector/cjdns/tree/patchset-v20.7},
with a synopsis in \ssref{syno-cjdns}; in the rest of this subsection,
we briefly introduce the build system, as well as our main
complaints with its original version.

\prog{cjdns}'s original build system was a set of \prog{Nodejs}
programs, with the main entry file \texttt{node\string_\lb{}build/make.js}
(used like \verb|node node_build/make.js|) calling into the core module
\texttt{node\string_build/\lb{}builder.js}; these two files added up to about
1700 lines, and when helper libraries and Shell wrappers were also taken into
account the sum rised to about 5000 lines (\cf\ commit \verb|e8d7367c|).
After refactoring, the build system has become a \verb|Makefile| and
a self-contained set of two \prog{Nodejs} scripts, \verb|jsmacro.js|
and \verb|jscfg.js|, the former used by both the latter and the
\verb|Makefile|; they add up to about 280 lines, and are used like
\verb|node jscfg.js && make|.  Other than the obvious burden of learning that
arised from the size of the original codebase, a heavy use of callback-based
concurrency to implement parallel build was also a source of headache.

\begin{figure}[htbp]\centering
\includegraphics[width = 0.8\textwidth]{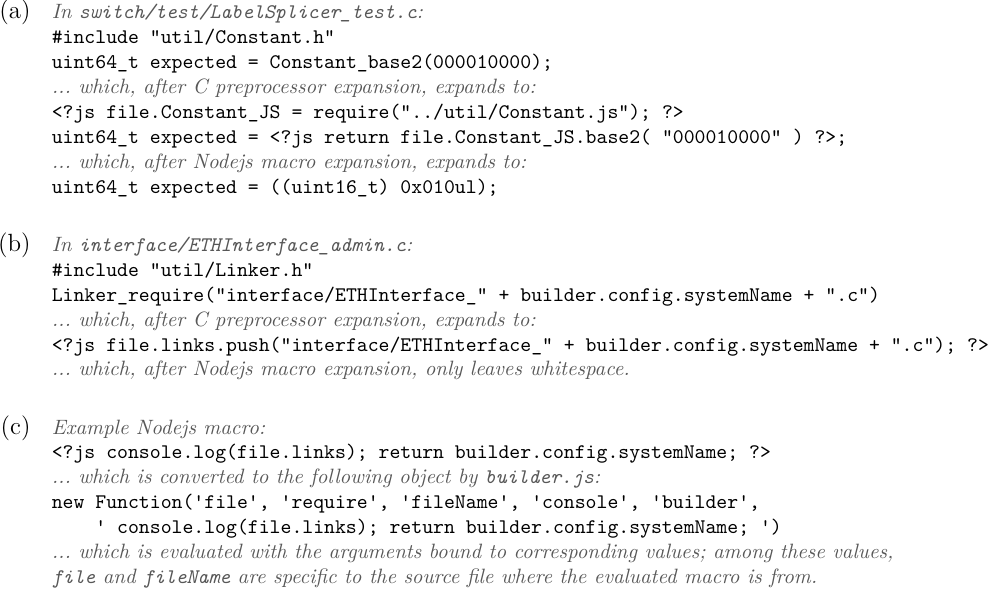}
\caption{%
	Use of \prog{Nodejs}-based macros (a) without or (b) with side effects
	between source files in \prog{cjdns}, and (c) these macros' access to the
	internal state of the builder object in \texttt{builder.js}; the examples
	shown are based on revisions \texttt{55410050}--\texttt{bd2ec9a5}.%
}
\label{fig:cjdns-macros0}
\end{figure}

But an even more serious issue we found was the additional complexity
in the internal state of the monolithic builder object created from
\verb|builder.js|.  \prog{cjdns}'s C source code uses \prog{Nodejs}-based macros
(\figref{cjdns-macros0}) in addition to regular macros, which in the original
codebase had unrestricted access to state attributes in the builder object;
this can be compared to the abuse of global variables in programming, and the
abuse of root privilege in system administration.  This lack of modularisation,
when combined with the already complicated state for other functionalities of
\verb|builder.js|, made it very hard to manually infer the exact steps to
build \prog{cjdns}.  Apart from the difficulty in extraction of build commands,
we also had no guarantee of the types of interactions, whether on purpose
or in error, between the \prog{Nodejs}-based macro expansion of different
source files: \eg\ the macro in \figref{cjdns-macros0}(b) has a side effect
that records the dependency of the current file on another file,
which the macro in \figref{cjdns-macros0}(a) does not have.

\subsection{Synopsis of refactoring}\label{ssec:syno-cjdns}

Our refactoring was based on \prog{cjdns} v20.7
(revision \verb|d832e269|), and consists of 15 commits:
\begin{itemize}
	\item \verb|55410050|: preparation to help check equivalence
		between code before and after each commit (\cf\ \ssref{int-repr}).
	\item \verb|21819aa7|--\verb|e5af6c56| (5 commits):
		simplification of the \prog{Nodejs}-based macro interface.
	\item \verb|bd2ec9a5|--\verb|d156a243| (4 commits):
		an incomplete migration to the simplified build system.
	\item \verb|1a6a1677|--\verb|6929bb03| (5 commits):
		removal of bundled dependencies, and completion of the migration.
\end{itemize}

As was suggested in \ssref{bg-cjdns}, the product of our refactoring is a
\prog{make}-based build system, which still uses \prog{Nodejs} in preparation
(comparable to the \verb|./configure| phase in more conventional build systems)
and expansion of macros; the resulting codebase, including the \verb|Makefile|
and the two \verb|.js| files, is intentionally minimised to reduce the
possibility of bugs and the burden of maintenance.  By migrating to \prog{make},
the build commands are made explicit; with the elimination of \verb|builder.js|,
both the ``callback hell'' and the complex state therein disappear;
the \prog{make}-induced macro expansion is done in a separate process
for each source file, effectively limiting undesired side effects between
files.  But in order to achieve the result above, we needed to classify the
\prog{Nodejs}-based macros to make sure those with desired side effects are
still in effect after refactoring.  After preliminary observation of the macros
involving the \verb|builder| variable (\figref{cjdns-macros0}(c)), we concluded
that inter-file side effects are only used to specify dependencies between
files.  This conclusion is non-trivial, as several of these macros involved
complex inline code around \verb|file.links| (\cf\ \verb|util/Setuid.h|),
or called into other \prog{Nodejs} scripts which manipulated \verb|builder|
directly (\cf\ \verb|util/Seccomp.h|).  In order to facilitate later refactoring
steps and potential code review (of the build system or its refactoring)
by others, commits \verb|21819aa7|--\verb|e5af6c56| were done to make this
property explicit: any code that did not operate on \verb|file.links| directly
or with the \verb|Linker_require| wrapper was moved into \verb|make.js|.

\begin{figure}[htbp]\centering
\includegraphics[width = 0.77\textwidth]{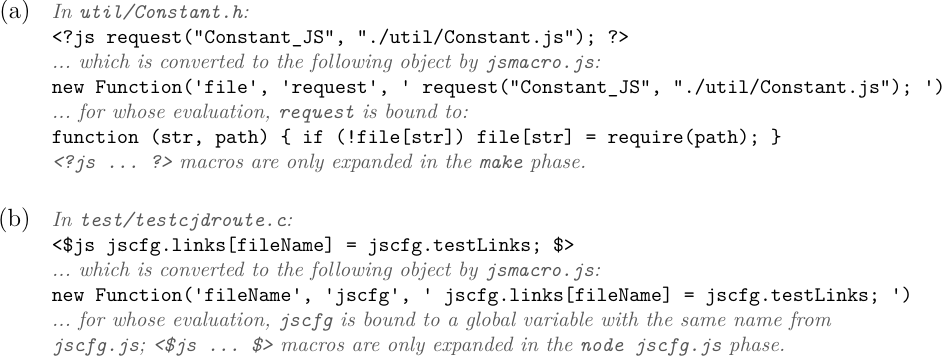}
\caption{The interface for \prog{Nodejs}-based macros after refactoring}
\label{fig:cjdns-macros1}
\end{figure}

After the previous commits, it became obvious that those \prog{Nodejs}-based
macros with inter-file side effects and those without do not depend on each
other, so we may separate their expansion into two phases: one to compute file
dependencies, and the other to do expansions without inter-file side effects;
they also naturally match our later separation between \verb|node jscfg.js|
and \verb|make|.  Again in order to make this separation explicit and prevent
any interaction between the two types of macros, we introduced and migrated to
the delimiters ``\verb|<$js ... $>|'' for dependency-related macros (which we
note are, after refactoring, mostly equivalent to loader \verb|#pragma|s in
the Plan 9 operating system \cite{pike2007}), while still using upstream's
``\verb|<?js ... ?>|'' for the rest (\figref{cjdns-macros1}).
With this change, full access to the builder object was
no longer needed, so most obstacles were cleared for the
migration to \verb|make| and the elimination of \verb|builder.js|.

The changes above constituted commits \verb|79314ba1|--\verb|d156a243|;
prior to them, the cosmetic commit \verb|bd2ec9a5| was done to eliminate
all nested \prog{Nodejs}-based macros, because we found that while the macro
expander is greatly simplified, the un-nested macros are not much longer
(or harder to read) than their nested counterparts.  The migration to
\verb|make| was incomplete after the previous commits, as the build system
still needed \verb|make.js| to build bundled dependencies \prog{libuv} and
\prog{NaCl}.  So with commits \verb|1a6a1677|--\verb|615daa90|, we migrated
to standalone (system-provided) versions of these dependencies, and finally
got rid of the old build system.  We note that commits \verb|2dcc7414|,
\verb|615daa90| and \verb|6929bb03|, which removed the old build system
or obsolete dependencies, can be regarded as direct applications
of the \emph{dead-code elimination} from compilers.

\subsection{Using intermediate files to check equivalence}\label{ssec:int-repr}

In \ssref{syno-cjdns}, what we did before commit \verb|d156a243| was
essentially transforming the old build system into \emph{equivalent systems with
increasingly rigorous constraints} on \prog{Nodejs}-based macros, with the final
constraint equivalent to that of the new build system.  However, the handling of
macros is quite intricate, and can often result in bugs; consequently, we should
not only rely on manual inspection and reasoning, but also find other ways
to check the equivalence between the build systems before and after each commit.
As has been mentioned, inter-file side effects of \prog{Nodejs}-based macros are
only used in the computation of file dependencies, so their correctness may be
checked by simply running the build process as missing dependencies would lead
to linker errors.  By comparing intermediate files with \prog{Nodejs}-based
macros expanded (\verb|.i| files generated by the build system), the correctness
of these macros can also be checked; their intra-file side effects, \eg\ the
binding of \verb|file.Constant_JS| in \figref{cjdns-macros0}(a), are also
checked since they are used to influence the expanded macros after all.

\begin{figure}[htbp]\centering
\includegraphics[width = 0.71\textwidth]{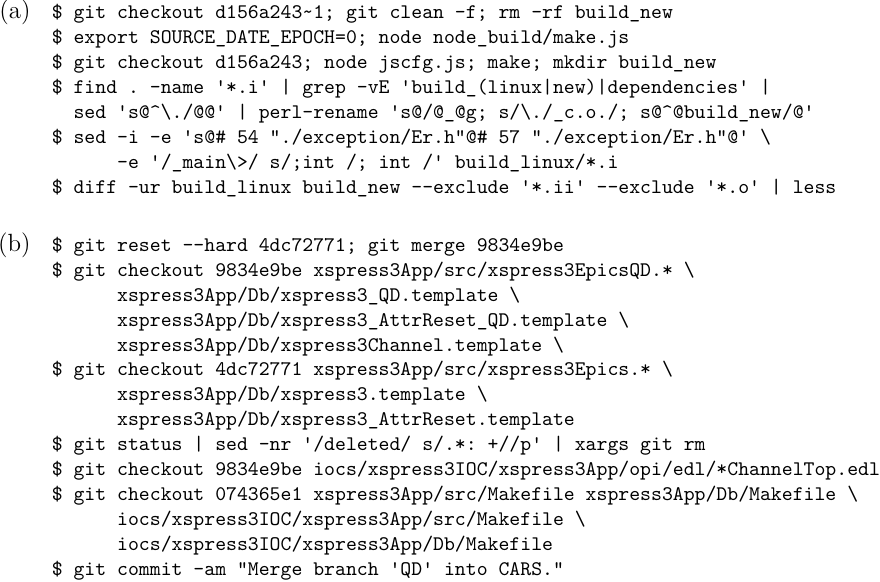}
\caption{%
	(a) Commands to compare the \texttt{.i} intermediate files before and after
	commit \texttt{d156a243} in \secref{cjdns} on one of our machines.  (b)
	Commands to merge the \texttt{QD} and \texttt{CARS} branches of Xspress3's
	IOC, at revisions \texttt{9834e9be} and \texttt{4dc72771} respectively.%
}
\label{fig:comp-merge}
\end{figure}

Some factors in the codebase could interfere with the kind of comparison: most
importantly the use of \prog{Nodejs}-generated random numbers in identifier
names, and the difference between the old and new build systems in the use of
relative paths (\verb|./path/to/file| \vs\ \verb|path/to/file|).  This issue was
resolved by commit \verb|55410050| and by using a fixed seed for random numbers
(supported in \verb|util/Constant.js| by setting the environment variable
\verb|${SOURCE_DATE_EPOCH}|).  With these changes, and with the assistance of
automation tools to systematically compress irrelevant differences, we were
able to reduce the difference in \verb|.i| files across each commit in the
\verb|21819aa7|--\verb|d156a243| range to an amount that can be confidently
checked by manual inspection: \eg\ the difference made by \verb|d156a243|,
the most non-trivial commit among them, can be checked with the commands
in \figref{comp-merge}(a).  Apart from the tools used to actually modify
files or directory trees, we also find the options \verb|--ignore-space-change|,
\verb|--ignore-all-space|, \verb|--ignore-blank-lines| of \verb|diff|
and \verb|--irreversible-delete| of \verb|git diff| highly helpful
in compressing differences in many of our refactoring projects.

\section{Refactoring the \prog{EPICS} IOC for Xspress3}\label{sec:xsp3}
\subsection{Background}

\begin{figure}[htbp]\centering
\includegraphics[width = 0.98\textwidth]{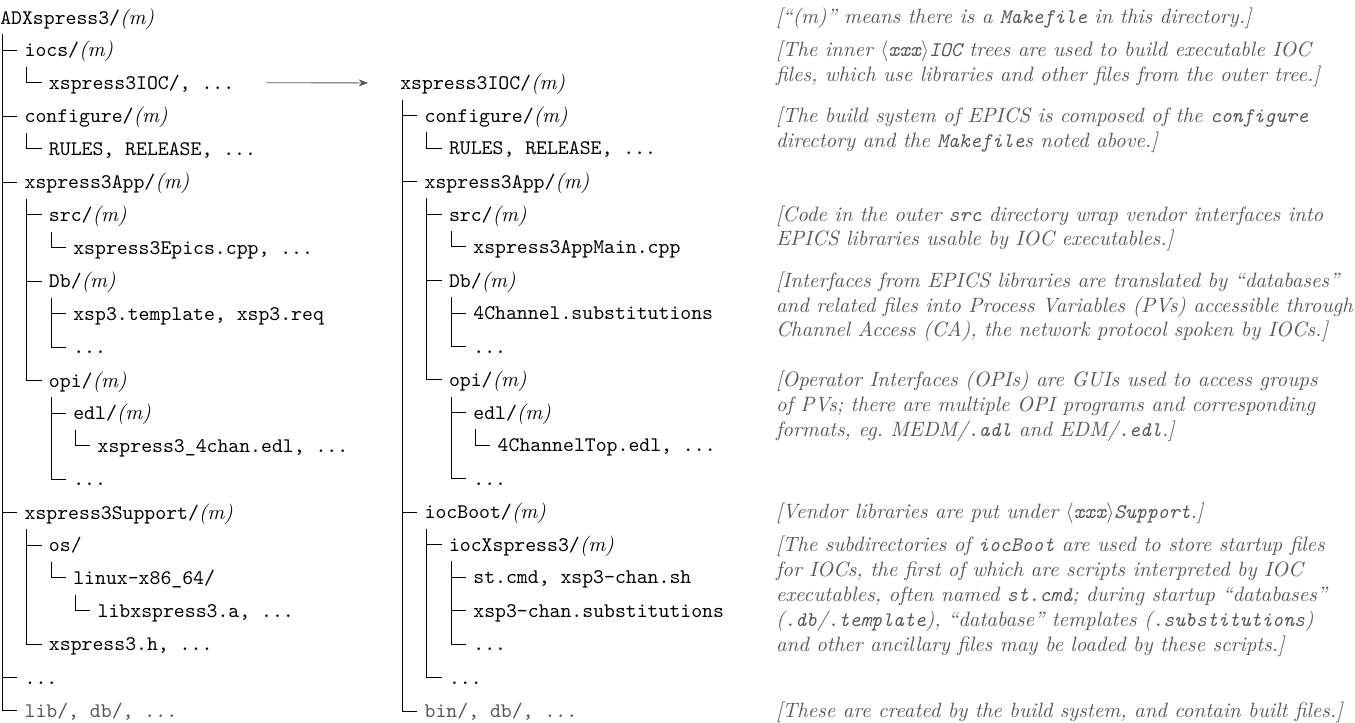}
\caption{Source code directory layout of our refactored Xspress3 IOC}
\label{fig:epics-tree}
\end{figure}

Xspress3 is a high-performance readout system manufactured by Quantum Detectors
for silicon drift detectors, which are widely used in X-ray fluorescence (XRF)
and X-ray absorption fine structure (XAFS) experiments.  An Xspress3 box has a
number of input channels, through each of which it generates an energy-resolved
histogram of X-ray photon counts.  A computer and one or more (which we denote
with \verb|${XSP3CARDS}|) Xspress3 boxes are connected in a subnet, where the
spectra can be read on the computer using Xspress3's libraries: assuming
the total number of channels is \verb|${XSP3CHANS}|, and the regular number
of histogram bins -- 4096 -- is used, each data frame will be a 2D array
of \verb|${XSP3CHANS}| $\times$ 4096 counts, the innermost dimension being
an 1D histogram.  At the Beijing Synchrotron Radiation Facility (BSRF)
and the High Energy Photon Source (HEPS), \prog{Experimental Physics
and Industrial Control System} (\prog{EPICS}) is widely used for
the control of beamline devices and the acquisition of data from them.
The control programs that talk to devices in \prog{EPICS} are
called Input-Output Controllers (IOCs); the source code directory
layout of a typical \prog{EPICS} IOC (especially when using its
\prog{areaDetector} framework) is outlined in \figref{epics-tree}.

Xspress3's IOC has two upstream branches, the Quantum Detector (QD) branch
(\url{https://github.com/quantumdetectors/xspress3-epics}) and the Center for
Advanced Radiation Sources (CARS, where main developers of \prog{areaDetector}
are) branch (\url{https://github.com/epics-modules/xspress3}).  Both of them
are derived from a common version with source files for different combinations
of \verb|${XSP3CHANS}| and \verb|${XSP3CARDS}| generated under the \verb|iocs|
directory by Diamond Light Source (DLS, which QD is close-related to)'s
\prog{iocbuilder} \cite{abbott2011}.  The QD branch still uses
\prog{iocbuilder}, which is several thousands lines of code even when
discounting addons; and although the subdirectories of \verb|iocs| are already
present in the \prog{Git} repository, each of them is again at least a few
thousands lines of code.  Additionally, the QD branch depends on an outdated
version of \prog{areaDetector}, and furtherly deviates from the latter
in a significant amount of DLS-specific customisation in its build system.
All these factors combined render the QD branch highly non-trivial to build,
and migration to up-to-date \prog{areaDetector} (with a more standardised
build system) seems to be difficult even for upstream developers, let alone
newcomers which additionally have to understand the huge codebase.

The CARS branch has only one directory under \verb|iocs|, but multiple
directories under \verb|iocBoot|, and the number of (\verb|${XSP3CHANS}|,
\verb|${XSP3CARDS}|) combinations available is much less than that on the QD
branch.  Moreover, unlike the QD branch where source files are more like just
batch-generated, there is a significant amount of manual modification specific
to some subdirectories of \verb|iocBoot|: \eg\ there are \verb|ioc_4Channel|,
\verb|iocxspress3-4Channel|, \verb|iocDualMini_4Channel| and
\verb|iocGSECARS-4Channel|; a newcomer cannot tell the difference between them.
With our refactored IOC (\url{https://github.com/CasperVector/ADXspress3},
with a synopsis of commits in \ssref{syno-xsp3}; what \figref{epics-tree} shows
is this IOC), the difference between (\verb|${XSP3CHANS}|, \verb|${XSP3CARDS}|)
combinations is minimised: the user only needs to specify these two variables
in \verb|st.cmd| under the unified \verb|iocXspress3| directory, and use
the code generator \verb|xsp3-chan.sh| (which is less than 30 lines
of self-contained code) to generate some ancillary files.

In addition to the minimisation of complexity in configuration, the contents
in \verb|iocXspress3| are, excluding example ancillary files generated by
\verb|xsp3-chan.sh|, around 200 lines.  During refactoring, our IOC has
been migrated to latest \prog{areaDetector}, with the build system completely
standardised so that it is built like regular \prog{areaDetector} IOCs (which
even the CARS branch is not).  Features specific to each upstream branch
have been merged, so that users do not need to worry about feature loss;
furthermore, backward compatibility with the upstreams is provided by
subdirectories of \verb|iocBoot| like \verb|iocXsp3QD| and \verb|iocXsp3CARS|;
we even fixed some bugs previously unnoticed by upstream developers.
The refactored IOC has been used by real beamline users in production at
BSRF, and works as expected; its refactoring is the subject of this section.
Although an expertise in \prog{EPICS} would facilitate understanding of
the discussion, we do not assume knowledge of \prog{EPICS} other than this
introduction (and especially \figref{epics-tree}).  Instead, we will show
how equivalence properties can help to ensure correctness by allowing
the programmer to consider the unrelated parts as big black boxes,
which are unchanged under the corresponding transformations.

\subsection{Synopsis of refactoring}\label{ssec:syno-xsp3}

Our refactoring was based on revision \verb|3d697abf| on the QD branch and
revision \verb|cc511b1e| on the CARS branch.  Changes were first committed
to the \prog{Git} branches \verb|QD| and \verb|CARS| in our repository to
make them converge, and then a \verb|git merge| was done.  There were 31 and 15
commits on branches \verb|QD| and \verb|CARS|, respectively, before the merger:
\begin{itemize}
	\item \verb|e7df7d1f|--\verb|a20f08de| on \verb|QD| (16 commits) and
		\verb|1a84b3e6|--\verb|4e51fd64| on \verb|CARS| (5 commits):
		merging peripheral files by cherry-picking commits from the other branch.
	\item \verb|55a04b53|--\verb|73ec6d50| on \verb|QD| (9 commits) and
		\verb|1cd766bb|--\verb|d701ca10| on \verb|CARS| (2 commits):
		early deduplication and migration of legacy dependencies
		for the QD branch, then standardisation for both branches.
	\item \verb|12326c5c|--\verb|9834e9be| on \verb|QD| (6 commits) and
		\verb|cd3106af|--\verb|4dc72771| on \verb|CARS| (8 commits):
		deep deduplication, introduction of \verb|iocXspress3|,
		preparation for the merger.
\end{itemize}
There were 9 more commits after the merger (commit \verb|074365e1|):
\begin{itemize}
	\item \verb|f67313a8|--\verb|a045eef5| (2 commits):
		removal of \prog{iocbuilder} remnants, update of documentation.
	\item \verb|64b0409e|--\verb|e2951c41| (4 commits):
		fixes for bugs in upstream code found in tests.
	\item \verb|bd2ec9a5|--\verb|59eecc35| (3 commits):
		completion of \verb|iocXspress3|, which previously lacked some
		upstream features for the sake of simplicity before the merger.
\end{itemize}
After these commits, our branch has been on par with the upstreams in terms of
features, and can be maintained independently.  Further commits are maintenance
changes or ports of new features from the upstreams, and are not discussed here.

We started by comparing the \verb|QD| and \verb|CARS| branches with \verb|diff|,
with its options \verb|-q| or \verb|-u| used as necessary; the differences
between the two branches were then reduced gradually, leaving only the minimum
needed for backward compatibility.  Among the differences between the two
branches, the most complicated were those involving files/directories inside
\verb|iocs| and the outer \verb|xspress3App/Db|; so in order to facilitate
further comparison between the two branches, we found it natural to first
reduce the differences in the rest of the codebase.  This was mainly done
by \verb|git log| operations on differing files/directories, followed by
\verb|git cherry-pick| operations to apply corresponding commits from the
other branch.  After this, we needed to choose a branch to mainly focus on;
we chose \verb|QD| as it was more uniform, which implied better susceptibility
to compression with automation tools, and perhaps better accumulation of
knowledge about the code for us during the procedure.  After the easier early
deduplication steps, we also migrated legacy dependency \prog{spectraPlugins}
for the \verb|QD| branch to \prog{NDPluginAttribute}, its successor available
from newer \prog{areaDetector}.  Now we were able to do closer merge operations
on peripheral files inside \verb|iocs|: not only the versions on the two
branches, but also those from other \prog{areaDetector} IOCs so that the build
system got maximally standardised; for the latter, \prog{ADSimDetector},
the IOC for a purely simulated area detector, was used as a main reference.

After the relatively easy steps above, the main differences left between the
two branches were those in files under \verb|iocBoot| and \verb|xspress3App/Db|,
which required deduplication of the code inside \verb|st.cmd| files anyway; for
this step, we again chose the more uniform \verb|QD| branch as our first target.
Commit \verb|5dc7acde| (and similarly, commits \verb|103d1585|--\verb|72166d3f|
on \verb|CARS|) was done to facilitate comparison operations on the command line
(as well as unified organisation of the startup files), then the inevitable
hard work of reading and understanding \verb|st.cmd| and \verb|.template| files
was done, resulting in the introduction of \verb|iocXspress3| in commits
\verb|f8f8abaf|--\verb|fe46d741| (and similarly, commits \verb|e6f40103|--%
\verb|b5554333| on \verb|CARS|).  With the changes above, the differences
between the two branches had almost been reduced to a minimum; then after a
small amount of preparation, a \verb|git merge| operation was done in commit
\verb|074365e1| (reproducible with the commands in \figref{comp-merge}(b);
as an afterthought, these commands should have been recorded in the
commit message).  With the knowledge acquired from the previous steps,
we did a final cleanup, unified the documentation, fixed bugs
revealed by simple test runs, and completed features missing
in \verb|iocXspress3|, thus concluding our refactoring.

\subsection{Methods to help ensure equivalence}\label{ssec:xsp3-methods}

The refactoring in \secref{cjdns} was most importantly based on the gradual
reduction of the differences between the old and new build systems in their
handling of \prog{Nodejs}-based macros; aside from using build tests to detect
errors, we also heavily depended on intermediate representations (IRs) to check
the equivalence across each commit.  In our refactoring of Xspress3's IOC,
apart from gradual unification (between \verb|QD| and \verb|CARS|, perhaps
as well as between them and \prog{ADSimDetector}), an equally important part
was deduplication: unification facilitated later deduplication by reducing
distraction, while deduplication facilitated later unification by compressing
repetitive code; as the codebase became increasingly uniform and succinct,
more and more was learned about it, culminating in the formation and completion
of \verb|iocXspress3|.  But unlike \secref{cjdns}, in our refactoring of
Xspress3's IOC, while build tests and test runs were still used, we could
not rely on IRs to check equivalence.  Instead, for the unification steps,
we mainly depended on experience, including both prior experience about
\prog{EPICS} and knowledge acquired in previous refactoring steps;
for the deduplication steps, we employed semiformal reasoning
extensively.  In this subsection, we mainly discuss the latter.

Like commits \verb|2dcc7414|, \verb|615daa90| and \verb|6929bb03| from
\secref{cjdns}, commits in the refactoring of Xspress3's IOC that removed
unnecessary files are essentially applications of the dead-code elimination;
commit \verb|3b315750| on \verb|CARS|, which removed unnecessary functions
and variables from the C++ source code (inside the outer \verb|src| directory
in \figref{epics-tree}), is a dead-code elimination under the classical
definition.  They are supported by the simple idea that a codebase with some
unnecessary code removed is equivalent to the original codebase, provided that
the removed code is truly unnecessary, both as reference for programmers and
as information source for tools like \prog{pydoc}.  When doing deduplication,
\emph{renaming} is perhaps the most frequently used code transformation, which
essentially substitutes all references to an old name with those to a new name:
\eg\ commits \verb|12326c5c| and \verb|cd3106af| which migrated from the
macro name \verb|${XSPRESS3}| to \verb|${ADXSPRESS3}|, as the latter is
more conventional in \prog{areaDetector}.  An important type of renaming is
\emph{relocation}, which renames files/directories: \eg\ commit \verb|0c619b81|
on \verb|QD| which switched to unified \verb|xspress3App| directories under
all subdirectories of \verb|iocs|, in preparation for the merger of these
subdirectories into \verb|xspress3IOC| in the next commit, \verb|4cceb49a|.

The equivalence across a renaming operation depends on the condition that
substituted names are exactly those which semantically match what we mean to
substitute: no false positive or false negative is allowed.  For codebases with
strong uniformity, it is advisable to automate renaming operations with tools
for text processing and file/directory manipulation, essentially creating code
transformers based on these tools; for most automated commits in this paper,
the transformers are trivial enough that their original implementations were
not recorded in the commit messages.  Unlike generic applications like code
uglifiers which have to consider corner cases like dynamically named variables
in possible inputs, in refactoring we usually only need to consider the
codebase in question, so false positives and false negatives in automated
pattern matching can be ruled out without using complex mechanisms like IRs:
\eg\ in commits \verb|12326c5c| and \verb|cd3106af|, the correct ranges for
text substitution were determined by running \verb|grep -r '\<XSPRESS3\>'|.
As another example, we note that in \ssref{syno-cjdns}, we equated direct
access to the builder object by \prog{Nodejs}-based macros with access to
the \verb|builder| variable; this is correct only because we had confirmed,
in advance with tools like \verb|grep|, that none of these macros accessed
the builder object indirectly, \eg\ through the \verb|file| variable.

As has been noted in \ssref{syno-xsp3}, even with every effort done to unify the
codebase and deduplicate the directory trees, the formation and completion of
\verb|iocXspress3| was still non-trivial.  It is because of the \emph{inherent
complexity} in the startup files, which was not much reduced in previous
simplification steps, although latter did help to minimise distraction.
This may be regard as a limitation of a purely equivalence-based approach:
while equivalence properties allow the programmer to consider unrelated parts
as black boxes, the inherent complexity in these black boxes is untouched.
In our attempts to simplify Xspress3's IOC, we also considered the unification
of the C++ source code on both branches, but eventually gave up because
of cost-to-effect ratio concerns resulted from the accumulated inherent
difference; in comparison, although the cherry-picking in the beginning
of our refactoring resulted in many commits, the accumulated difference was
much smaller.  We also note that for IOCs internally developed at or heavily
customised for BSRF/HEPS, we do perform refactoring/rewriting on the C++ level
(an example can be seen at \url{https://github.com/CasperVector/ADPandABlocks}),
as we consider it a part of their development.

Nevertheless, after closer inspection of the commit histories of the C++
source code on both branches (especially those after their common ancestor,
\verb|ed3fbdd4|), we noticed that for each commit on the QD branch (\eg\ %
\verb|91b4bd2f|), there is a commit on the CARS branch (\eg\ \verb|ef44fa8a|)
implementing the same functionality.  If we consider this as some kind of
\emph{equivalence with approximation}, we may say changes to the C++ source
code on the QD branch are a subset of corresponding changes on the CARS
branch, therefore in our README file it is suggested that the QD variant
of the IOC executable be only used with the QD startup files for backward
compatibility.  However, we recently noticed that a modification only present
on the CARS branch resulted in performance regression, as it switched from
block-by-block readout to frame-by-frame readout, leading to much bigger
per-frame overhead; for this reason, at BSRF and HEPS we also use the QD
IOC executable with \verb|iocXspress3|.  This kind of issues are admittedly
difficult to spot with a purely equivalence-based approach, as they
require the programmer to delve into the internals of the black boxes.

\section{Discussion and conclusion}
\subsection{Discussion}\label{ssec:discuss}

In our simplification of the codebases in question, we attempted to approximate
some kind of complexity lower-bounds in the resulting products, which we feel
is a natural match with our methodology.  However, it should also be obvious
that once an original codebase and a final target have been chosen, the
complexity lower-bound is determined for the set of possible refactoring paths.
Therefore even if the value of a target might be disputed, the techniques in
this paper would still be helpful in estimating the practicality of the target.

As can be seen from Subsection \ref{ssec:syno-cjdns} and \ref{ssec:syno-xsp3},
we consider refactoring as an iterative procedure to the programmer, where
internal structures in the codebase are gradually discovered after prior
simplifications; for this reason, we consider refactoring as a great way
of learning.  But in order to keep the list of code transformations
(here represented as \prog{Git} commits) manageable, we often perform
\verb|git rebase| operations to merge commits that are related but not
necessarily adjacent in time.  So the order of commits usually do not strictly
correlate to the order of iterative learning: \eg\ the changes used in
\ssref{int-repr} were merged into commit \verb|55410050|, but the necessity
of changes to \verb|.js| files in this commit was not revealed until an early
version of commit \verb|d156a243|.  Similarly, in \ref{ssec:syno-cjdns}, the
\verb|jscfg| object (\figref{cjdns-macros1}(b)) appeared in the (rebased)
commit history much earlier than the real migration to two-phase macros.

As with other kinds of programming, in refactoring it is possible to introduce
bugs.  In the transformation-based approach in this paper, the refactoring
steps (\prog{Git} commits) may be seen as loosely coupled modules with clear
semantics, so it is logical to merge fixes to bugs found in early testing
(\ie\ usually not after the refactored codebase has been deployed in production)
with the earlier commits they fix, using \verb|git rebase|.  Counting this kind
of bugfixes and the merging of other logically related commits, we estimate that
in a typical refactoring project done in our way, the number of commits that
actually occurred could be a few times the length of the rebased commit history
left in the \prog{Git} repository.  So necessity of the rebasing operations
can be easily seen: otherwise the commit history would be flooded with
bugfixes, conflicting with our goal to keep it manageable.

Among the commits \verb|21819aa7|--\verb|e5af6c56| in \secref{cjdns},
\verb|21819aa7| and \verb|4a919ff6| were not strictly related to the
simplification around \verb|file.links| inside some macros.  The former was
done to move all related conditional logics around \verb|Linker_require|
into one header, to make it easier for the programmer to reason about them;
the latter would have felt more natural along with commits \verb|bd2ec9a5|%
--\verb|d156a243|, but was done so that obstacles for the next commit,
\verb|e5af6c56|, could be cleared.  We also note that while commit
\verb|21819aa7| is, essentially, in large part an \emph{inlining}
transformation, the inverse of inling transformations can also
be seen in commits \verb|21819aa7|--\verb|e5af6c56|,
as has been suggested in \ssref{syno-cjdns}.

As can be seen from \secref{xsp3}, uniformity facilitates refactoring by
helping the programmer to concentrate on the essential differences between
code to be merged.  Similarly, clear separation between rigid applications
and their variable configurations facilitates deployment, which is why
we introduced \verb|iocXspress3|.  Proper standardisation facilitates
development, which is why we emphasise the standardisation of the build
system for \prog{EPICS} IOCs, whether open-source or developed internally
at BSRF/HEPS \cite{liu2022}.  We also note that we have refactored too many
internally developed IOCs, often derived from bloated templates generated
by \verb|makeBaseApp.pl| from \prog{EPICS}, that instead of repeating small
transformations like dead-code elimination for each IOC, now we usually extract
files of interest from the given IOCs directly and put them into clean pre-made
templates. From this, we consider experience as a kind of proof for equivalence
with approximation (\cf\ \ssref{xsp3-methods}); it can be used to theoretically
explain, \eg, why the migration to standalone \prog{libuv} and \prog{NaCl} in
\secref{cjdns} did not cause problems despite the large amount of code change.

A main objective for our refactoring of \emph{ADPandABlocks} (mentioned
in \secref{xsp3}) was to make convert it to a pure data-readout IOC, as the
extra functionalities have been reimplemented elsewhere at BSRF/HEPS in a
much smaller codebase.  The most important step in this refactoring is commit
\verb|05aeb612|, which shows a pattern frequently seen by us in dead-code
elimination for refactoring: the previous commit, \verb|16e02887|, removed
some \emph{EPICS} ``records'' deemed unnecessary by us; they not only had
(obverse) dependencies in the original codebase, but also inverse dependencies,
the latter being ``hidden'' dead code in essence.  To correctly remove
these code, we analysed the dependence between them, and found the real
root of the obsolete subgraph in the dependency graph -- the function
\verb|pollCommandPortC()| in the original C++ source code; after that
we became able to confidently do the ordinary dead-code elimination.

\subsection{Conclusion}

By treating the procedure of software refactoring as composing code
transformations, and compressing repetitive transformations with
automation tools, we can often obtain representations of refactoring
processes short enough that their correctness can be analysed manually.
Unlike in compilers, in refactoring we usually only need to consider the
codebase in question, so regular text processing can be extensively used,
fully exploiting patterns only present in the codebase.  Aside from
the direct application of code transformations from compilers, like the
dead-code elimination and the inlining transformation (along with its
inverse transformation), many other kinds of equivalence properties may also
be exploited.  In this paper, mainly demonstrated are the equivalence across
renaming/relocation operations, the approximated equivalence supported by
experience and the approximated equivalence between commits on different
\prog{Git} branches doing similar things.  Two refactoring projects are
given as the main examples in this paper: refactoring of \prog{cjdns}'s build
system, which was based on transformation between equivalent systems with
increasingly rigorous constraints; refactoring of Xspress3's \prog{EPICS} IOC,
which was based on an alternation between unification and deduplication steps.

\section*{Acknowledgements}

The author would like to thank both the Unix and Lisp communities for
the invaluable inspirations from them, without which the ideas behind
this paper would not have come into existence.  The author would also
like to thank the developers of the \prog{EPICS} IOC (both the QD and
CARS branches) for Xspress3, whose design decisions (including both
good and bad ones) were the direct motivations for this paper.  This
work was supported by the Technological Innovation Program of Institute
of High Energy Physics of Chinese Academy of Sciences (E25455U210).

\bibliography{art5}

\begin{thebibliography}{5}
\providecommand{\natexlab}[1]{#1}
\providecommand{\url}[1]{\texttt{#1}}
\expandafter\ifx\csname urlstyle\endcsname\relax
  \providecommand{\doi}[1]{doi: #1}\else
  \providecommand{\doi}{doi: \begingroup \urlstyle{rm}\Url}\fi

\bibitem[Keep(2012)]{keep2012}
A.~W. Keep.
\newblock \emph{A Nanopass Framework for Commercial Compiler Development}.
\newblock PhD thesis, School of Informatics and Computing, Indiana University,
  2012.

\bibitem[Bentley et~al.(1986)Bentley, Knuth, and McIlroy]{bentley1986}
J.~Bentley, D.~Knuth, and D.~McIlroy.
\newblock Programming pearls: A literate program.
\newblock \emph{Commun. ACM}, 29\penalty0 (6):\penalty0 471--483, 1986.

\bibitem[Pike(2007)]{pike2007}
R.~Pike.
\newblock How to use the plan 9 c compiler, 2007.
\newblock URL \url{https://plan9.io/sys/doc/comp.pdf}.

\bibitem[Abbott and Cobb(2011)]{abbott2011}
M.~G. Abbott and T.~Cobb.
\newblock An epics ioc builder.
\newblock In \emph{Proceedings of the 13th International Conference on
  Accelerators and Large Experimental Physics Control Systems (ICALEPCS2011)},
  number MOPMU032, pages 506--509, Grenoble, France, 2011.

\bibitem[Liu et~al.(2022)Liu, Dong, and Li]{liu2022}
Y.~Liu, X.-W. Dong, and G.~Li.
\newblock Better automation of beamline control at heps.
\newblock \emph{J. Synchrotron Rad.}, 29\penalty0 (3):\penalty0 687--697, 2022.

\end{thebibliography}
\end{document}